\documentclass[review,11pt]{elsarticle}

\usepackage{graphicx}
\usepackage{rotating}
\usepackage{subcaption}
\usepackage{amssymb,amsmath} 

\DeclareMathOperator*{\argmax}{arg\,max}

\newcommand{\nn}{\nonumber}
\newcommand{\mbf}[1]{\mbox{$\mathbf #1$}}    
\newcommand{\mref}[1]{(\ref{#1})}

\newcommand{\SNR}{\textrm{SNR}}
\newcommand{\JNR}{\textrm{JNR}}
\newcommand{\be}{\vspace*{-0.035cm}\begin{eqnarray}}
\newcommand{\ee}{\end{eqnarray}\vspace*{-0.035cm}}
\newcommand{\beIEEE}{\vspace*{-0.02cm}\begin{IEEEeqnarray}{rCl}}
\newcommand{\eeIEEE}{\end{IEEEeqnarray}\vspace*{-0.02cm}}

\journal{Signal Processing}

\begin{document}
\setlength{\interdisplaylinepenalty}{2500}

\begin{frontmatter}

\title{An Automated Window Selection Procedure For DFT Based Detection Schemes To Reduce Windowing SNR Loss}
\author[METU]{\c{C}a\u{g}atay Candan}
\ead{ccandan@metu.edu.tr}
\address[METU]{Department of Electrical and Electronics
Engineering, Middle East Technical University (METU), 06800,
Ankara, Turkey.}

\begin{abstract}
The classical spectrum analysis methods utilize window functions to reduce the masking effect of a strong spectral component over weaker components. The main cost of side-lobe reduction is the reduction of signal-to-noise ratio (SNR) level of the output spectrum. We present a single snapshot method which optimizes the selection of most suitable window function among a finite set of candidate windows, say rectangle, Hamming, Blackman windows, for each spectral bin. The main goal is to utilize different window functions at each spectral output depending on the interference level encountered at that spectral bin so as to reduce the SNR loss associated with the windowing operation. Stated differently, the windows with strong interference suppression capabilities are utilized only when a sufficiently powerful interferer is corrupting the spectral bin of interest is present, i.e. only when this window is needed. The achieved reduction in the windowing SNR loss can be important for the detection of low SNR targets.
\end{abstract}

\end{frontmatter}

\section{Introduction}
A common approach, if not the most common, in the spectral analysis of signals is the windowed Fourier transformation, which is the well known periodogram approach. In many applications, including signal-to-noise ratio (SNR) sensitive detection applications, a nominal window is pre-selected and applied irrespective of the operational SNR throughout the deployment of the detection system. Yet, it is known that the desired feature of window function (side-lobe suppression) comes at the cost of output SNR loss. As an example, the frequent choice of Hamming window results in an SNR loss of 1.35 dB, which can be important for the detection of targets at low SNR or equivalently for the extension of the instrumented range of a radar system. The main goal of this study is to pose the window selection problem as a hypothesis test and present an automated procedure that keeps the SNR loss due to windowing operation at a minimum. To do that, we assume that the system has a finite number of window functions at its disposal (say rectangle, Hamming, Blackman windows) and aims to select the most suitable window function for each spectral output, i.e.  for each discrete Fourier transformation (DFT) bin, through the Bayesian hypothesis testing. The proposed method is based on a single snapshot of data and is applicable in all conventional detection schemes utilizing windowed DFT operation.

\begin{figure}[t]
\centering
\includegraphics[width=12cm]{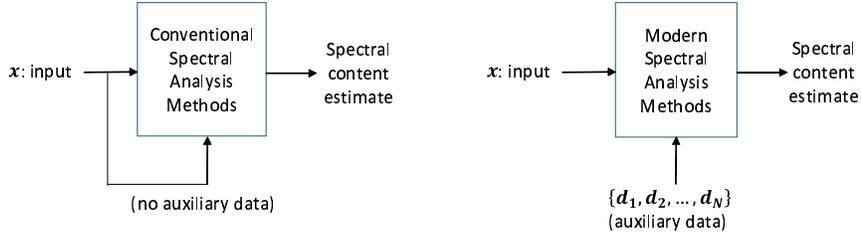}
\caption{Conventional and Modern Spectral Analysis Methods}
\label{figintro}
\end{figure}

The spectral analysis is a well established topic of statistical signal processing closely linked with several applications in speech processing, radar signal processing, remote sensing. The main goal of spectral analysis is to detect and accurately estimate the power of each spectral component of the input. The methods to this aim can be categorized as
single and multiple snapshot methods as illustrated in Figure~\ref{figintro}. The single snapshots methods are, in general,  data-independent methods based on the windowed Fourier transformation. Multiple snapshot methods, such as the Capon's method, are data-dependent methods and based on the estimation and minimization (ensemble) average value interference at the output. The multiple snapshot methods use more information, such as the interference/signal auto-correlation function and yield superior results, in general. In certain applications, a-priori information on the signal and interference statistics may not be available or the sensing scheme may not suitable for an ensemble characterization. For the conventional pulse-Doppler radar systems, a single snapshot vector, composed of slow-time samples from a range cell, is available  to detect the presence of a moving target in a range cell which can be contaminated with clutter, jammers and potentially other targets. The conventional processing chain typically includes a stage of windowed DFT. A suitable window, say Hamming window having a good  side-lobe suppression (43 dB for the Hamming window) is selected to reduce the shadowing effect of strong undesired component over the target component. Unfortunately, this choice brings an SNR loss, which is 1.35 dB for the Hamming window that can be compensated by increasing the transmit power. This requires a factor of $10^{(1.35/10)} \approx 1.36$ fold increase in the number of transmit elements of a phase array system operating at peak power limitation. SNR losses due to processing, called signal processing loss in the radar signal processing literature, should be avoided as much as possible; since their compound effect on the system design, say on the power budget, can be a significant factor affecting the monetary cost of the system.

The selection of a suitable window function is an application or scenario specific choice based on some trade-offs. As noted before, the main benefit of windowing is the reduction of the signal sidelobes in the output spectrum at the expense of widened mainlobe (resolution loss) and SNR loss, as documented in \cite[Ch.6]{poratbook}. The main goal of this study is to optimize the window selection such that the resulting average SNR loss due to windowing is negligibly small. To this aim, we propose to use a set of conventional windows, say rectangle, Hamming and a Chebyshev windows providing 13, 43, 120 dB side-lobe suppression ratios, pose the window selection as a hypothesis testing problem and apply the principles of the Bayesian hypothesis testing. Upon an analysis of the resulting hypothesis testing based method, we suggest a reduced complexity window selection method and examine its performance.

In the literature, the method known as the dual-apodization method (for two windows) and its extension to the multiple windows (multi-apodization) resembles the proposed line of study, \cite{stankwitz1995}. In dual-apodization, the DFT of the input is calculated twice with two different window functions. For a specific DFT bin, the dual apodization output is the output spectrum sample of two windows having the minimum magnitude. The motivation of dual-apodization method can be most easily seen for the noise-free operation. In the absence of noise, the window with the better side-lobe suppression is selected for the side-lobe bins by the mentioned calculation of minimum DFT magnitude. For the bins which are located in the main-lobe, the window with the smaller beamwidth, the higher resolution window, is selected. Hence, the inclusion of a non-linearity in the processing chain, jointly enables both high resolution and small side-lobes, which is not possible with linear processing methods. This method has found applications mainly in the imaging applications \cite{Munson2000,SVA2} where SNR is not the most major issue of concern as in the detection applications. To the best of our knowledge, the performance of dual-apodization is not examined from the detection theoretic point of view except the study of \cite{BurakBalciTez}. In this work, we present the comparison of the suggested hypothesis testing based window selection method with the multi-apodization method.

\section{Background}
\label{sec:background}
We consider the following signal model,
\be
r[n] = \underbrace{\sqrt{\gamma_s}e^{j(\omega_s n + \phi_s)}}_{s[n]} +
       \underbrace{\sqrt{\gamma_j}e^{j(\omega_j n + \phi_j)}}_{j[n]}
       + v[n], \quad n=\{0,1, \ldots , N-1\}.
\label{eq1}
\ee
Here $s[n]$ denotes the signal of interest, a complex exponential signal with frequency $\omega_s$; $j[n]$ denotes the intentional or unintentional jamming signal corrupting the observations $r[n]$; $v[n]$ is zero mean, unit variance
complex valued white noise with circularly symmetric Gaussian distribution, $v[n] \sim \textrm{CN}(0,1)$. The phase values of signal and jammer components, $\phi_s$ and $\phi_j$, are assumed to be independent random variables with a uniform distribution in $[0,2\pi)$. The parameters $\gamma_s$ and $\gamma_j$ are independent, exponentially distributed random variables with mean values $\bar{\gamma}_s$ and $\bar{\gamma}_j$, respectively. The input SNR and jammer-to-noise ratio ($\JNR$) is defined as $\SNR=\frac{E\{|s[n]|^2\}}{E\{|v[n]|^2\}} = \bar{\gamma}_s$,
$\JNR =\frac{E\{|j[n]|^2\}}{E\{|v[n]|^2\}} = \bar{\gamma}_j$. (The model given in \mref{eq1} corresponds to the Rayleigh faded target signal observed under Rayleigh faded rank-1 interference (jammer) and white Gaussian noise.)

Equation \mref{eq1} can be written in vector form as $\mbf{r} = \sqrt{\gamma_s}e^{j\phi_s}\mbf{s} + \sqrt{\gamma_j}e^{j\phi_j}\mbf{j} + \mbf{v}$, where $\mbf{s}$ is an $N \times 1 $ column vector with the entries $e^{j\omega_s n}$, $n=\{0,1, \ldots , N-1\}$. (The vector $\mbf{j}$ is defined similarly.) Throughout this work, we prefer to express the frequency variables, $\omega_s$ or $\omega_j$, with the units of DFT bins, $\omega_s = 2\pi f_s /N$, where $N$ is the number of observations. With this definition, $f_s$ becomes a real valued parameter in the interval $[0,N)$. This definition simplifies the description and perception of the numerical values associated with main-lobe width, the frequency difference between signals etc.

\begin{figure}[t]
\centering
\includegraphics[height=9cm]{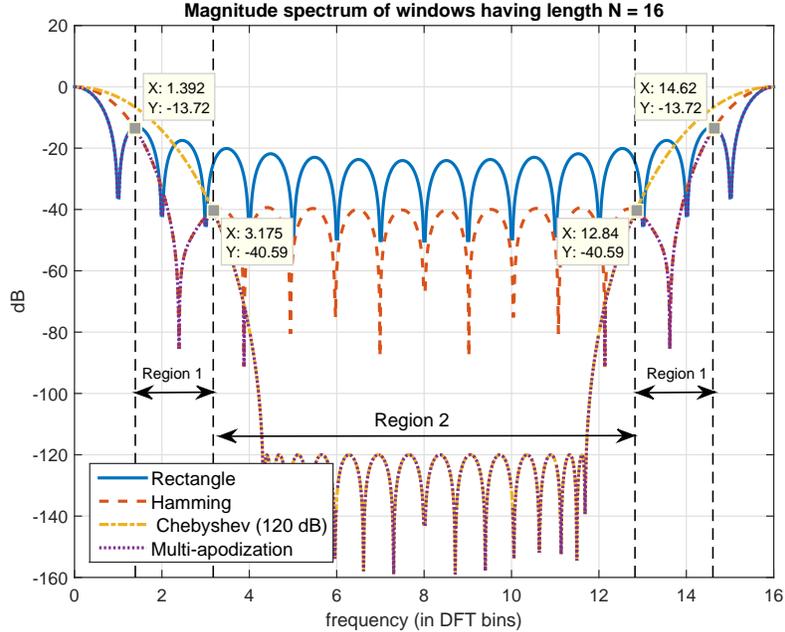}
\caption{Magnitude spectrum of Rectangle, Hamming, Chebyshev windows and the multi-apodization spectrum.}
\label{fig_windows}
\end{figure}

 Figure~\ref{fig_windows} shows the magnitude spectrum of rectangle, Hamming and Chebyshev windows of length 16. In the presence of white noise and absence of jamming, the optimal detector, maximizing the output SNR, is the detector matched to the signal vector $\mbf{w} = \mbf{s}$ and the optimal decision statistics for the detection application is $|\mbf{w}^H \mbf{r}|^2$, \cite{richards_frsp}. This corresponds to the processing of the input data with the rectangle window.  When jamming is present, SNR maximizing filter is the whitened matched filter detector, that is $\mbf{w} = (\mbf{I} + \gamma_j \mbf{j}\mbf{j}^H)^{-1}\mbf{s}$ and the decision statistics is $|\mbf{s}^H(\mbf{I} + \gamma_j \mbf{j}\mbf{j}^H)^{-1}\mbf{r}|^2$. In this work, we assume that the user does not have the capacity to implement the optimal filter due to the lack of statistical information on the jammer. Instead, user selects the most suitable window from a set of candidate windows to reduce the effect of jamming signal. The decision statistics becomes $|\mbf{s}^H \mbf{D}_\textrm{win} \mbf{r}|^2$ where $\mbf{D}_\textrm{win}$ is a diagonal matrix whose diagonal entries are the samples of the window function selected.

To understand the factors effecting the window choice, assume that the target is located at the DC bin. For this target, a very strong jammer in the Region 2 of Figure~\ref{fig_windows} requires the application of Chebyshev window. Yet, the Hamming window, which has a better SNR loss than Chebyshev window, can be sufficient for a weaker jammer in the same region. On the other hand, even for very strong jammers in Region 1 of Figure~\ref{fig_windows}, the Chebyshev window is not suitable, since there is no side-lobe suppression due to large main-lobe widening of this window. Hence for Region 1 jammers, the possible windows of choice are limited to rectangle or Hamming window.  In this study, our goal is to automate such reasoning procedures and present a window selection method with a negligible SNR loss.

In Figure~\ref{fig_windows}, the multi-apodization output of three windows is also illustrated. For this case, the multi-apodization output magnitude is simply
\be
X_{MA}(e^{j\omega}) = \min\{|X_R(e^{j\omega})|, |X_H(e^{j\omega})|, |X_{Ch}(e^{j\omega})|\}
\label{eq2}
\ee
where $X_R(\cdot), X_R(\cdot), X_{Ch}(\cdot)$ is the spectrum of rectangle, Hamming and Chebyshev windows and $X_{MA}(e^{j\omega})$ is the multi-apodization output, \cite{Munson2000,BurakBalciTez}. It can be seen from \mref{eq2}
that the multi-apodization output is simply the selection of the magnitude spectrum with the smallest magnitude for each spectral component, \cite{stankwitz1995}. In the absence of noise, the multi-apodization yields jointly high resolution (main-lobe width of rectangle window) and large side-lobe suppression, as shown in  Figure~\ref{fig_windows}. For the noisy scenarios, the performance of multi-apodization, in terms of SNR loss, is not immediately clear and a topic of investigation in this study. The multi-apodization idea has found some applications in the synthetic aperture imaging (SAR) applications, \cite{SVA2,EffectiveSVA} where SNR loss is not the main concern. The situation is quite different in the pulse-doppler radar systems where much fewer observations (slow time samples) are available for the target detection. An insightful explanation for the multi-apodization method given in \cite{EffectiveSVA} states that multi-apodization is, in principle, equivalent to the single snapshot version of Capon's method. In this work, our goal can be more ambitiously stated as to develop a better alternative for the open title of single snapshot Capon's method.

\section{Proposed Approach}
The proposed approach is described with the window functions illustrated in Figure~\ref{fig_windows}. Without any loss of generality, we may assume that the signal of interest is at DC bin and the jamming  power is localized either in Region 1 or Region 2 of Figure~\ref{fig_windows}.


Our goal is to determine the jammer activity through M-ary hypothesis testing and apply a suitable window based on the jamming activity level. The k'th hypothesis can be written as follows:
\be
\textrm{H}_k : \quad \mbf{r} = \sqrt{\gamma_s}e^{j\phi_s}\mbf{s} + \sqrt{\gamma_{j_k}}e^{j\phi_j}\mbf{j} + \mbf{v}, \quad k=\{1,\ldots,M\}.
\ee
The vectors $\mbf{s}, \mbf{j}, \mbf{v}$ refer to signal, jammer and noise vectors as defined in Section~\ref{sec:background}. The $\gamma_s$ is nuisance parameter of the test appearing in all hypotheses. The critical parameter of the k'th hypothesis is the JNR, $\bar{\gamma}_{j_k}$. We assume that
$\bar{\gamma}_{j_1} < \bar{\gamma}_{j_2} < \ldots < \bar{\gamma}_{j_M}$, that is
the weakest jammer case is represented with $H_1$.

For any observation vector $\mbf{r}$, the index of the hypothesis with the highest a-posteriori probability can be expressed as
\be
\widehat{k} = \argmax_{1\leq k \leq M} P(H_k | \mbf{r}).
\label{eq3}
\ee
We treat the selected hypothesis as an indicator of the jamming activity level and associate a window function for each hypothesis. Note that \mref{eq3} can also be written in terms of the likelihood ratios as follows,
\be
\widehat{k} = \argmax_{1\leq k \leq M}
\frac{ P(H_k | \mbf{r}) }
{ P(H_1 | \mbf{r}) }
= \argmax_{1\leq k \leq M}
\frac{ P(H_k)}{P(H_1)}
\frac{ f_\mathbf{R}( \mbf{r} | H_k ) }{ f_\mathbf{R}( \mbf{r} | H_1 ) }
\label{eq4},
\ee
where $P(H_k)$ is the a-priori probability of the k'th hypothesis and the rightmost term is the likelihood ratio of the $k$th and first hypothesis. In many scenarios, the a-priori probabilities of hypothesis are unknown and are taken as $P(H_k) = 1/M$ for $k=\{1,2,\ldots,M\}$. Then, $\widehat{k}$ reduces to $\argmax_{1\leq k \leq M} \frac{ f_\mathbf{R}( \mbf{r} | H_k ) }{ f_\mathbf{R}( \mbf{r} | H_1 ) }$.

\subsection{Window Selection Test Based on the Likelihood Ratio}
\label{test_lr}
With the definitions given Section~\ref{sec:background}, $f_\mathbf{R}(\mathbf{r} | H_k)$ is Gaussian density with zero mean and covariance matrix $\mathbf{R}_k = \bar{\gamma}_{s}\mathbf{ss}^H +
\bar{\gamma}_{j_k}\mathbf{R}_j^n
+\mathbf{I}$, where $\mathbf{R}_j^n$ is the normalized jammer covariance matrix with unit elements on its diagonal. The product of normalized jammer covariance matrix and average jammer power for the k'th hypothesis is denoted as $\mathbf{R}_{j_k} = \bar{\gamma}_{j_k}\mathbf{R}_j^n$.

We express the eigendecomposition of $\mathbf{R}_{j_k}$ as $\mathbf{R}_{j_k} = \mathbf{E}_j\mathbf{\Lambda}_{j_k}\mathbf{E}_j^H$. Here, the columns of $\mathbf{E}_j$ matrix are  the eigenvectors spanning the jamming space. The matrix $\mathbf{\Lambda}_{j_k}$ is a diagonal matrix having the eigenvalues $\mathbf{R}_{j_k}$ on its diagonal. The eigenvalues of $\mathbf{R}_{j_k}$ is $\bar{\gamma}_{j_k} \times \lambda_i^n$, that is the product of $\bar{\gamma}_{j_k}$ and the eigenvalues of normalized jammer covariance matrix $\mathbf{R}_j^n$.


The log-likelihood ratio of the k'th and first hypothesis can be written as:
\be
\log \left(
\frac{f_\mathbf{R}(\mathbf{r} | H_k)}
{f_\mathbf{R}(\mathbf{r} | H_1)}
\right)
 = \log|\mathbf{R}_1| - \log|\mathbf{R}_k|
 + \mathbf{r}^H (\mathbf{R}_1^{-1} - \mathbf{R}_k^{-1}) \mathbf{r}.
 \label{LRTeq5}
 \ee
Temporarily, we call jammer plus noise covariance matrix for the $k$th hypothesis as $\mathbf{R}_{j_k+n} =
 \bar{\gamma}_{j_k}\mathbf{R}_j +\mathbf{I} = \mathbf{R}_{j_k} + \mathbf{I}$, then $\mathbf{R}_k$ appearing in the likelihood ratio becomes $\mathbf{R}_k
 = \bar{\gamma}_{s}\mathbf{ss}^H + \mathbf{R}_{j_k + n}$. With this definition, the quadratic product in \mref{LRTeq5} reduces to
 \be
 \mathbf{r}^H \mathbf{R}_k^{-1} \mathbf{r}
 =
 \mathbf{r}^H \mathbf{R}_{j_k + n}^{-1} \mathbf{r}
 -
 \frac{1}{1 + \mathbf{s}^H \mathbf{R}_{j_k + n}^{-1} \mathbf{s} }
 | \mathbf{r}^H \mathbf{R}_{j_k + n}^{-1} \mathbf{s}  |^2.
 \label{eq6}
 \ee
 The jammer plus noise covariance matrix, $\mathbf{R}_{j_k + n} = \mathbf{R}_{j_k} + \mathbf{I}$, can be eigendecomposed as
 $\mathbf{R}_{j_k + n} = \mathbf{E}_j\mathbf{\Lambda}_{j_k + n}\mathbf{E}_j^H +
 \mathbf{E}_n \mathbf{E}_n^H$. It should be noted that jammer plus noise covariance matrix is a full rank matrix; therefore,  the columns of $\mathbf{E}_n$ matrix, whose span is the noise space, is orthogonal to the jammer space, the column space of $\mathbf{E}_j$ matrix.

Using the eigendecomposition of $\mathbf{R}_{j_k + n}$, we can express
 $ \mathbf{s}^H \mathbf{R}_{j_k + n}^{-1} \mathbf{s} $ term appearing in \mref{eq6} as follows:
 \be
 \label{eq7}
 \mathbf{s}^H \mathbf{R}_{j_k + n}^{-1} \mathbf{s} =
 \underbrace { \mathbf{s}^H \mathbf{E}_j\mathbf{\Lambda}_{j_k + n}^{-1}\mathbf{E}_j^H \mathbf{s} } _{\approx 0}
 +
 \mathbf{s}^H\mathbf{E}_n \mathbf{E}_n^H \mathbf{s}
 \approx |\mathbf{E}_n^H \mathbf{s}|^2.
  \ee
 Here, it is assumed that the signal vector has negligible power in the jammer space.
 We may consider that the jammer is in Region 2 of Figure~\ref{fig_windows}; therefore the projection of signal power to jammer space is negligible. (The leakage of signal power can be considered as negligible unless the SNR is extremely is high which is a case of little concern for the detection problems.)

Similarly, we can express the term $\mathbf{r}^H \mathbf{R}_{j_k + n}^{-1} \mathbf{s}$ in \mref{eq6} as
 \be
 \label{eq8}
 \mathbf{r}^H \mathbf{R}_{j_k + n}^{-1} \mathbf{s}
 =
  \underbrace { \mathbf{r}^H \mathbf{E}_j\mathbf{\Lambda}_{j_k + n}^{-1}\mathbf{E}_j^H \mathbf{s} } _{\approx 0}
 +
 \mathbf{r}^H\mathbf{E}_n \mathbf{E}_n^H \mathbf{s}
 \approx \mathbf{r}^H \mathbf{E}_n \mathbf{E}_n^H \mathbf{s}.
 \ee
Finally, the term $\mathbf{r}^H \mathbf{R}_{j_k + n}^{-1} \mathbf{r}=
\mathbf{r}^H  (\mathbf{R}_{j_k} + \mathbf{I})^{-1} \mathbf{r}^H  $ in \mref{eq6} becomes
 \be
   \label{eq9}
 \mathbf{r}^H \mathbf{R}_{j_k + n}^{-1} \mathbf{r}
&  = &
 \mathbf{r}^H \mathbf{E}_j\mathbf{\Lambda}_{j_k + n}^{-1}\mathbf{E}_j^H \mathbf{r}
 +
  \mathbf{r}^H\mathbf{E}_n \mathbf{E}_n^H \mathbf{r} \nn \\
& = &
  \sum_{i=1}^{N_j} \frac{1}{\bar{\gamma}_{j,k} \lambda_i^{n} + 1} |\mathbf{e}_i^H \mathbf{r}|^2
  + |\mathbf{E}_n^H \mathbf{r}|^2,
\ee
where $N_j$ is rank of $\mathbf{R}_{j_k} = \bar{\gamma}_{j_k}\mathbf{R}_j^n $ and $\mathbf{e}_i$ is the eigenvector of $\mathbf{R}_{j_k}$ with the eigenvalue $\bar{\gamma}_{j,k} \lambda_i^{n}$.

Using equations \mref{eq7}, \mref{eq8} and \mref{eq9}, we can simplify the quadratic term appearing on the left side of \mref{eq6}, $\mathbf{r}^H \mathbf{R}_k^{-1} \mathbf{r}$, as
 \be
&& \phantom{\hspace{-1cm}}
 \mathbf{r}^H \mathbf{R}_k^{-1} \mathbf{r}
  \approx \\
&& \phantom{\hspace{-1cm}}
 \mathbf{r}^H \mathbf{E}_j\mathbf{\Lambda}_{j_k + n}^{-1}\mathbf{E}_j^H \mathbf{r}
 +
  \mathbf{r}^H\mathbf{E}_n \mathbf{E}_n^H \mathbf{r}
 - \frac{1}{1 + |\mathbf{E}_n^H \mathbf{s}|^2}
 | \mathbf{r}^H \mathbf{E}_n \mathbf{E}_n^H \mathbf{s} |^2.
 \nn
 \ee
 This concludes the simplification of the quadratic term in the log-likelihood ratio expression in \mref{LRTeq5}. The remaining term to be simplified in this equation is the determinant of $\mathbf{R}_k = \bar{\gamma}_{s}\mathbf{ss}^H + \mathbf{R}_{j_k + n}$ matrix:
 \be
 \det(\mathbf{R}_k)
 & = &
 ( 1 + \bar{\gamma}_{s} \mathbf{s}^H \mathbf{R}_{j_k + n}^{-1} \mathbf{s} )
 \det( \mathbf{R}_{j_k + n})
  \nn \\
 & \stackrel{(a)}{=} &
 ( 1 + \bar{\gamma}_{s} |\mathbf{E}_n^H \mathbf{s}|^2 )
 \det( \mathbf{R}_{j_k + n})
   \nn \\
 & \stackrel{(b)}{=} &
 ( 1 + \bar{\gamma}_{s} |\mathbf{E}_n^H \mathbf{s}|^2 )
 \prod_{i=1}^{N}({\bar{\gamma}_{j,k} \lambda_i^{n} + 1})
\label{eq11}
   \ee
 In line-(a) of equation \mref{eq11}, the relation $ \mathbf{s}^H \mathbf{R}_{j_k + n}^{-1} \mathbf{s} \approx |\mathbf{E}_n^H \mathbf{s}|^2$ is utilized one more time (see equation \mref{eq7}). In line-(b), the determinant is computed via the product of the eigenvalues.

 With the substitution of relations given by equations \mref{eq9} and \mref{eq11} into the log-likelihood ratio given in \mref{LRTeq5}, we get the following expression,
\be
& & 
\phantom{\hspace{-1cm}}
\log \left(
\frac{f_\mathbf{R}(\mathbf{r} | H_k)}
{f_\mathbf{R}(\mathbf{r} | H_1)}
\right)
 \approx   \\ 
& & \phantom{\hspace{-1cm}}
\sum_{i=1}^{N_j}\log \left(
\frac{ \bar{\gamma}_{j,1} \lambda_i^{n} + 1}
{ \bar{\gamma}_{j,k} \lambda_i^{n} + 1  } \right)
+
\sum_{i=1}^{N_j}
\left(
\frac{1}{\bar{\gamma}_{j,1} \lambda_i^{n} + 1}
-
\frac{1}{\bar{\gamma}_{j,k} \lambda_i^{n} + 1}
\right)
|\mathbf{e}_i^H \mathbf{r}|^2. \nn
\label{aLRTeq12}
\ee
In the last expression $\mathbf{e}_i$ and $\lambda_i^n$ are the eigenvector of normalized jammer covariance matrix $\mathbf{R}_j^n$. Next, we need to construct a hypothesis for each candidate window function with a proper selection of hypothesis parameters.

\begin{figure}[t]
\centering
\includegraphics[width=10.5cm]{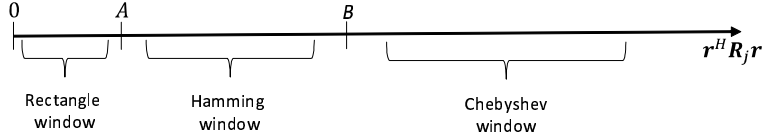}
\caption{An illustration for the partitioning of the real line into three parts.}
\label{figreal_line}
\end{figure}

\textbf{On the selection of $\mathbf{R}_j^n$ matrix:} We assume that the jammer frequency is assumed to be uniformly distributed in the side-lobe region of the window. As a concrete example, we may consider the side-lobe region of N=16 point Chebyshev window, shown as the Region 2 of Figure~\ref{fig_windows}. This region ranges from 3.175 to 12.84 DFT bin as can be seen from Figure~\ref{fig_windows}. (The same region with unit of radians per sample corresponds to $\Theta_1 = 2\pi/N \times 3.175$ and $\Theta_2 = 2\pi/N \times 12.84$, where $N=16$.)

The jammer with the frequency $\theta$ corresponds to $\mbf{j}_\theta = [1 \,\, e^{j\theta} \,\, e^{j2\theta} \,\, \ldots \,\, e^{j(N-1) \theta}]^T$. The jammer frequency is assumed to be uniformly distributed in the interval $[\theta_1,\theta_2]$, which is [3.175, 12.84] DFT bins for the Region 2 shown in Figure~\ref{fig_windows}. The auto-correlation matrix for the jammer is $E\{ \mbf{j}_\theta \mbf{j}_\theta^H \}$, where the expectation operator $E\{ \cdot \}$ is over the random variable corresponding to the frequency $\Theta$. The $m$th row, $n$th column entry of the matrix $\mathbf{R}_j^n$ can be expressed as:
\be
\left[ \mathbf{R}_j^n \right]_{mn}
& = &  
\frac{\theta_2 - \theta_1}{2\pi}
\left[ E_\Theta \{ \mbf{j}_\theta \mbf{j}_\theta^H \} \right]_{mn}
  =  \frac{1}{2\pi}
 \int_{\theta_1}^{\theta_2}
 { e^{j(m-n)\theta} d\theta  } \nn \\
 & = &
\left\{
\begin{tabular}{ll}
$(\theta_2 - \theta_1)/2\pi$, & $m=n$ \\
$\frac{\exp({j(m-n)\theta_2}) - \exp({j(m-n)\theta_1})}{j2\pi(m-n)}$, & $m \neq n$
\end{tabular}.
\right.
\label{eq13}
\ee


\textbf{On the selection of $\bar{\gamma}_{j_k}$ parameter:} The candidate windows should have significantly different side-lobe suppression ratios. The rectangle, Hamming and Chebyshev windows of Figure~\ref{fig_windows} have the peak side-lobe suppression ratios of 13, 40 and 120 dB. For each window, we suggest to use half the peak side-lobe ratio in dB as $\bar{\gamma}_{j_k}$ parameter. For the mentioned case, this results in $\bar{\gamma}_{j_1} = 6.5$ dB, $\bar{\gamma}_{j_2} = 20$ dB and $\bar{\gamma}_{j_3} = 60$ dB. This choice stems from the consideration that for a jammer of 15 dB JNR, we may almost equally prefer to use either Hamming or rectangle window, but not the Chebyshev window.

\subsection{A Simple Window Selection Test Based On Likelihood Ratio Approximation}
We present a simpler test based on an approximation to the log-likelihood ratio given in \mref{aLRTeq12}. From the definition of normalized jammer covariance matrix given by \mref{eq13}, it can be verified that the eigenvalues of this matrix, denoted as $\lambda_i^{n}$, are clustered around 0 and 1. (This fact can be checked by noticing that $\mathbf{R}_j^n$ matrix given in \mref{eq13} is the similarity transform of the discrete prolate spheroidal sequence (DPSS) generating matrix having the $m$th row, $n$th column entries $\sin(\theta_2(m-n))/(\pi(m-n)$ whose eigenvalues are known to have a sharp transition between 1 and 0,  \cite[p.213]{papoulisSA}.) By substituting $\lambda_i^{n} \approx 1$ in \mref{aLRTeq12}, we get the following expression
\be
&& \phantom{\hspace{-1cm}}  
\log \left(
\frac{f_\mathbf{R}(\mathbf{r} | H_k)}
{f_\mathbf{R}(\mathbf{r} | H_1)}
\right) \approx   \nn \\
&& \phantom{\hspace{-1cm}}
N_j
\log \left(
\frac{ \bar{\gamma}_{j,1} + 1}
{ \bar{\gamma}_{j,k} + 1 }
\right)
+
\left(
\frac{1}{\bar{\gamma}_{j,1} + 1}
-
\frac{1}{\bar{\gamma}_{j,k} + 1}
\right)
\sum_{i=1}^{N_j}
|\mathbf{e}_i^H \mathbf{r}|^2,
\nn \\
& & \phantom{\hspace{-1cm}}  =
\alpha_k+ \beta_k
\mathbf{r}^H \mathbf{R}_j^n \mathbf{r}.
\label{eq14}
\ee
where $\sum_{i=1}^{N_j}
|\mathbf{e}_i^H \mathbf{r}|^2 = \mathbf{r}^H \mathbf{R}_j^n \mathbf{r}$ is an estimate of the jammer power.  Hence, the log-likelihood ratio reduces to an affine function of $\mathbf{r}^H \mathbf{R}_j^n \mathbf{r}$, $\alpha_k + \beta_k \mathbf{r}^H \mathbf{R}_j^n \mathbf{r}$. Hence, the association of window functions to the hypotheses can be simply done by partitioning the real line to M disjoint sets as shown in Figure~\ref{figreal_line}. The decision boundaries, the points $A$ and $B$, in Figure~\ref{figreal_line} are the JNR levels for the window switching indicating the boundaries between low, medium, high JNR levels.

\textbf{On the determination of decision boundaries:} We assume that the window functions are ordered in the increasing order of interference suppressing capabilities. (For the case shown in Figure~\ref{fig_windows}, the order is rectangle, Hamming and Chebyshev windows, respectively.) The goal is to sequentially set the decision boundaries, that is to determine the boundary for the rectangle and Hamming windows first (point A in Figure~\ref{figreal_line}) and then the boundary for the Hamming and Chebyshev window (point B in Figure~\ref{figreal_line}). It can be said that for a candidate window set with $M$ windows, we need to determine $M-1$ boundaries by pairwise comparing consecutive windows ordered in the increasing order of PSL.

\begin{figure}[t!]
    \centering
    \begin{subfigure}[t]{0.5\textwidth}
        \centering
\includegraphics[height=7cm]{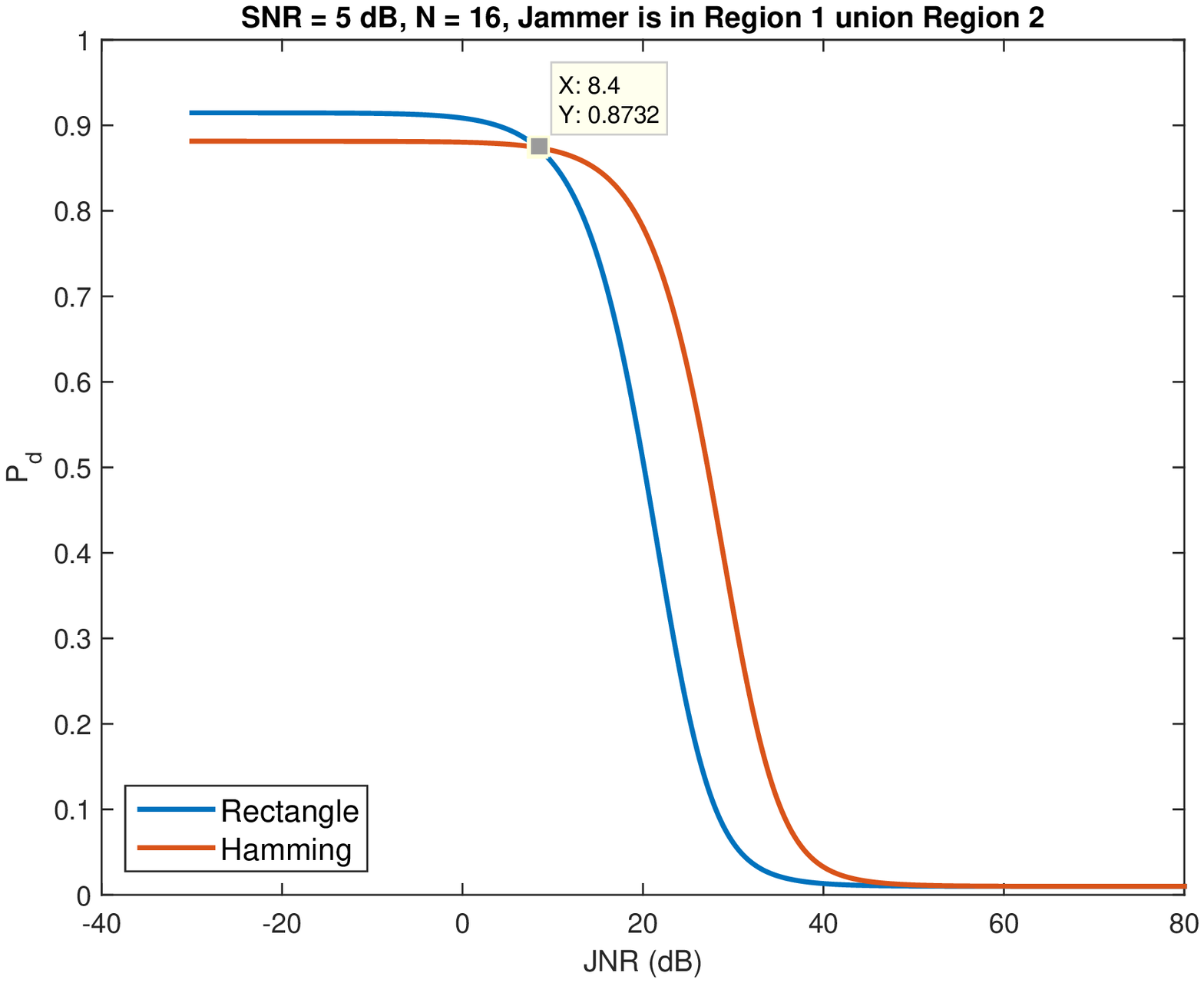}
        \caption{Rectangle vs. Hamming}
    \end{subfigure}%
    \begin{subfigure}[t]{0.5\textwidth}
        \centering
\includegraphics[height=7cm]{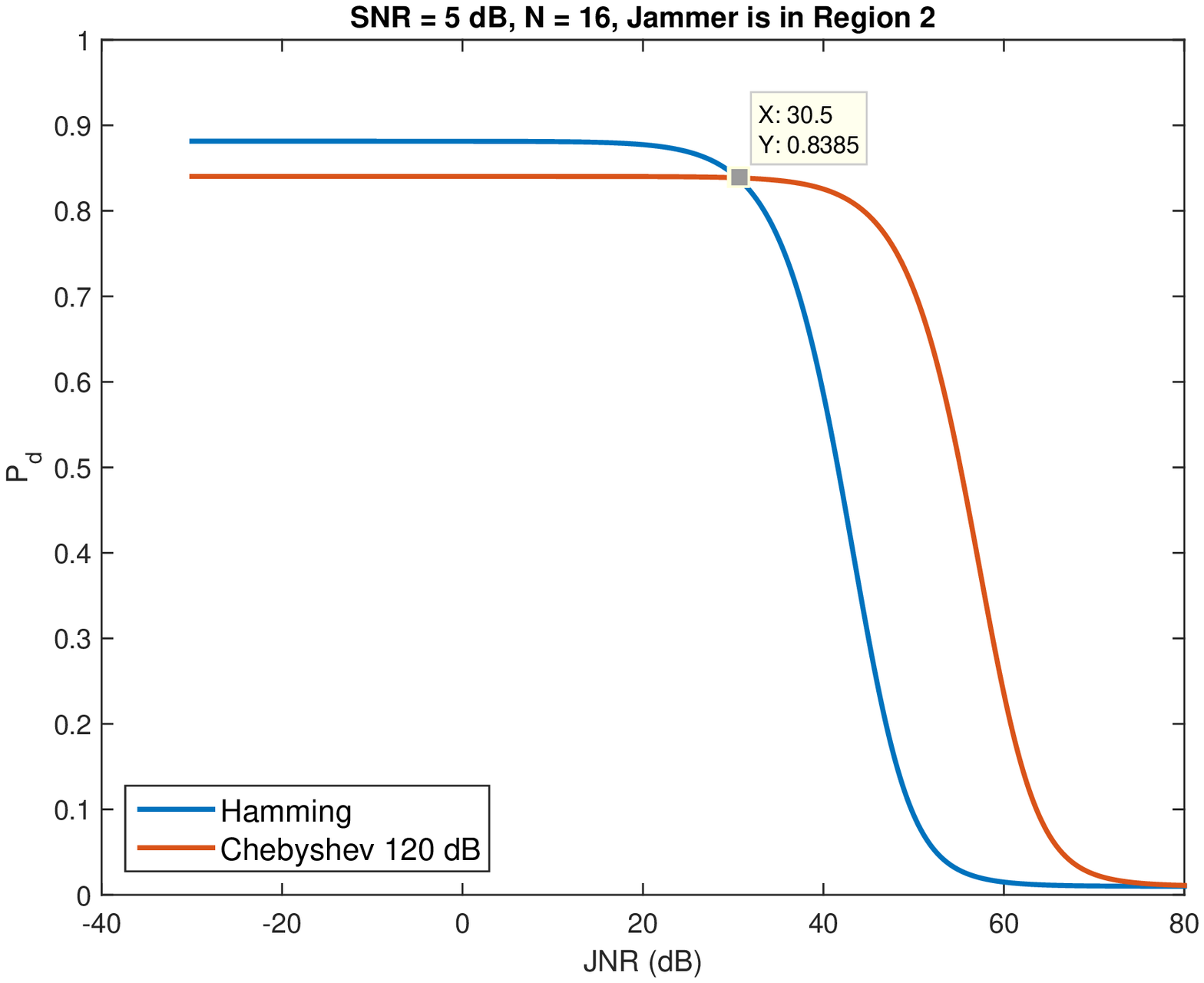}
        \caption{Hamming vs. Chebyshev (120 dB)}
    \end{subfigure}
    \caption{The receiver operating curves (ROC) for windowed DFT detectors, $P_{FA} = 10^{-2}$. }
\label{fig2}
\end{figure}

Figure~\ref{fig2}a shows the receiver operating curve (ROC) for the rectangle and Hamming windowed processed received vector $\mathbf{r}$ whose signal model is given in \mref{eq1}. Since the signal, interference and noise are assumed to be jointly Gaussian distributed, the detection problem reduces to the problem known as the detection of Gaussian signals in Gaussian noise,  \cite[Ch.9]{vantreesp3}. The relation between probability of detection and probability of false alarm, as depicted in Figure~\ref{fig2} is $P_d = P_{FA}^{1/(1 + \textrm{SJNR})}$, where $\textrm{SJNR}$ refers to the signal-to-jammer-plus-noise-ratio. SJNR at the windowed DFT detector output is written as follows:
\be
\textrm{SJNR}_k
= \frac { \bar{\gamma}_{s} |\mathbf{w}^H_k \mathbf{s}|^2 }
{\bar{\gamma}_{j} \mathbf{w}_k^H \mathbf{R}_j^n \mathbf{w}_k + \|\mathbf{w}_k\|^2}.
\ee
Here $\mathbf{w}_k$ is the $k$th window and without any loss of generality, the vector $\mbf{s}$ can be taken as the vector of all ones, that is the signal can be assumed at the DC bin. (If the signal is not at the DC bin, $\mbf{w}_k$ should be window function modulated to the frequency of the signal.)

From the ROC curve given in \mref{fig2}a, it can be concluded that when JNR is less than 8.4 dB, which is the intersection point of ROC curves for two windows, the rectangle window is a better choice yielding higher probability of detection value. The intersection point of ROC curves for two windows can be easily calculated by equating $\textrm{SJNR}$ values of two windows. If the windows are indexed as 1 and 2, the relation $\textrm{SJNR}_1 = \textrm{SJNR}_2$, yields the JNR value for the intersection point as
\be
\bar{\gamma}_{j} =
\frac{
|\mathbf{w}^H_1 \mathbf{s}|^2 \|\mathbf{w}_2\|^2
-
|\mathbf{w}^H_2 \mathbf{s}|^2 \|\mathbf{w}_1\|^2
}
{
|\mathbf{w}^H_2 \mathbf{s}|^2
\mathbf{w}^H_1 \mathbf{R}_j^n \mathbf{w}_1
-
|\mathbf{w}^H_1 \mathbf{s}|^2
\mathbf{w}^H_2 \mathbf{R}_j^n \mathbf{w}_2.
}
\label{eq16}
\ee
In \mref{eq16}, the matrix $\mathbf{R}_j^n$ is the normalized jammer covariance matrix defined in \mref{eq13}. The parameters $\theta_1$ and $\theta_2$ of this matrix is determined by considering the side-lobe suppression regions of both windows. As shown in Figure\ref{fig_windows}, the side-lobe suppression region for both rectangle and Hamming windows range from 1.392 to 14.62 in terms of DFT bins. It should be noted that the decision boundaries given by \mref{eq16} is invariant to SNR and probability of false alarm. Hence, the boundary point can be calculated once and can be utilized for all ROC curves.

As a summary, the window selection on the approximate log-likelihood relation is based on the quantization of the metric  $d = \mathbf{r}^H \mathbf{R}_j^n \mathbf{r}/N$. For the case shown in Figure~\ref{fig_windows}, if the $d$ value is smaller than 8.4 dB, then rectangle window is selected; for $d$ values in between $8.4$ and $30.5$ Hamming window is selected and for $d > 30.5$ results in the selection of Chebyshev window.

In the development of the test based on the likelihood ratio (given in Section \ref{test_lr}) and its approximate version (given in this section), the interference is assumed to be in the suppression band of all windows, which is the part of spectrum shown as Region 2 of Figure~\ref{fig_windows}. If the interference lies in the transition band of a particular window (shown as Region 1 of Figure~\ref{fig_windows}), then that window should be discarded. As an example, for the interference lying in Region 1 of Figure~\ref{fig_windows}, the Chebyshev window should not be utilized at all; since the  Chebyshev window does not provide any improvement in interference suppression in comparison to the Hamming window for this jammer. On the contrary, the Chebyshev window amplifies the interference power in comparison to the Hamming window. Hence, in spite of its superior side-lobe suppression capacity for the Region 2 interferers, the Chebyshev window should be avoided for the Region 1 jammers.

We integrate the idea of disabling some of the windows based on the detected jammer region to the existing scheme and present the suggested method in Table~\ref{table_algorithm}. For further clarification, we also present a ready-to-use MATLAB code with application examples in \cite{ccandan-code}.

\begin{sidewaystable}[t]
\caption{The suggested window selection method. Inputs are $\mathbf{r}$: $N \times 1 $ input vector, $L$ window functions, $\alpha$: DFT bin index of interest. Output is $k^{select}$ where
$k^{select}$ is the index of the window function to be used for the spectral analysis of $\alpha$'th DFT bin. (A Matlab implementation is available at \cite{ccandan-code}.}
\begin{tabular}{|l|p{15cm}|}
\hline
\parbox[t]{0.9cm}{Step 0:\\ (Init.)}
 &
 a) Order $L$ window functions in the increasing order sidelobe suppression ratio. (Example: Rectangle, Hamming, Chebyshev windows should be ordered as the first, second and third windows, respectively.)

    b) Calculate the $L-1$ decision boundaries for $L$ windows, denoted as $\bar{\gamma}_{j}[k]$, by setting $\mbf{w}_1$ as the $k$th window function and $\mbf{w}_2$ as the $(k+1)$th window function, $k=\{1,2, \ldots, L-1\}$, in \mref{eq16}. Also set $\bar{\gamma}_{j}[0]=0$.

    c) Determine the stop-band of each window with the index $k\geq 2$. (Example: In figure~\ref{fig_windows}, the stop-band for the Hamming and Chebyshev windows are 1.392 to 14.62 DFT bins and 3.175 to 12.84 DFT bins, respectively.)

    d) To select the window for the output DFT bin, $\alpha$'th DFT bin, $0 \leq \alpha \leq N-1$, modulate the input vector $\mathbf{r}$ to the zero'th DFT bin, i.e.
    $\mathbf{r} \leftarrow \mathbf{D}_\alpha \mathbf{r}$ where $\mathbf{D}_\alpha$
    is a diagonal matrix with the diagonal entries $\exp(-j\frac{2\pi \alpha n}{N})$,
    $0 \leq n \leq N-1$.
\\
\hline
Step 1: & For the k'th window with the stop-band $\theta_1$ to $\theta_2$, calculate $\mathbf{M}_k = \mathbf{R}_j^n$, $2 \leq k\leq L$ from \mref{eq13}.
\\
\hline
Step 2: & Calculate $d_k = \mathbf{r}^H \mathbf{M}_k \mathbf{r}/N$ for  $2 \leq k\leq L$, where $\mbf{r}$ is the modulated input vector of dimensions $N \times 1$.
\\
\hline
Step 3: & For $k=\{2, 3, \ldots, L-1\}$, check whether $d_{k} > 2d_{k+1}$ condition is satisfied or not. If the condition is not satisfied for any $k$, set $k^{\mathrm{max}} = L$; else set $k^{\mathrm{max}}$ as the lowest $k^{\mathrm{max}}$ value for which the condition is satisfied.  (window disabling)
\\
\hline
Step 4: & Return $k^{\mathrm{select}}$ such that $\bar{\gamma}_{j}[k^{\mathrm{select}}-1] \leq d_2 \leq \bar{\gamma}_{j}[k^{\mathrm{select}}]$.
, where $1 \leq k^{\mathrm{select}} \leq k^{\mathrm{max}}$.
\\
\hline
\end{tabular}
\label{table_algorithm}
\end{sidewaystable}

The suggested algorithm can be briefly explained as follows: Step 0 is the initialization step where the stop bands of the windows, the decision boundaries are calculated. In Step 1, the normalized covariance matrix for an interferer lying in the stop-band of each window is calculated. In Step 2, the jamming power residing in the stop band of each window is calculated. Step 3 implements the idea of window disabling based on the jammer spectral location. Step 4 is the quantization of estimated jamming power for the selection of the window function, as illustrated in Figure~\ref{figreal_line}.

\section{Numerical Comparisons}
We present a set of numerical comparisons of the suggested method with the conventional windowed DFT detectors and multi-apodization method.

\textbf{Case 1: Interference in the suppression band of all windows:} To ease the presentation, we continue with the detectors utilizing the rectangle, Hamming and Chebyshev windows whose magnitude spectrum is illustrated in Figure~\ref{fig_windows}. For this case, it is assumed that the interference lies in Region 2 of Figure~\ref{fig_windows}. More specifically, the interference lies in the band where all three windows have the ability of interference suppression at different capacities.

\begin{figure}[t!]
    \centering
    \begin{subfigure}[t]{0.5\textwidth}
        \centering
\includegraphics[height=6.5cm]{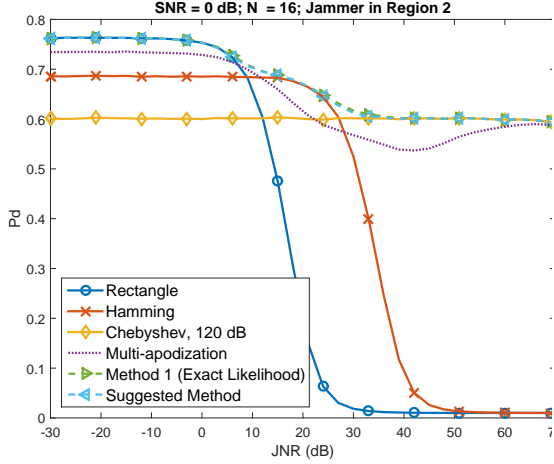}
        \caption{Pd vs. JNR}
    \end{subfigure}%
    \begin{subfigure}[t]{0.5\textwidth}
        \centering
\includegraphics[height=6.5cm]{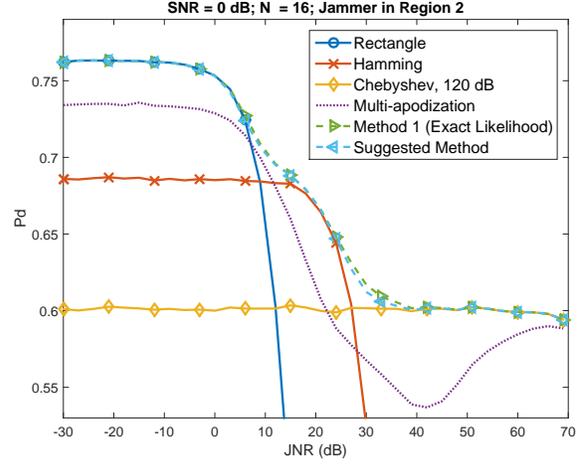}
        \caption{Pd vs. JNR (zoomed)}
    \end{subfigure}

\begin{subfigure}[t]{0.5\textwidth}
        \centering
\includegraphics[height=6.5cm]{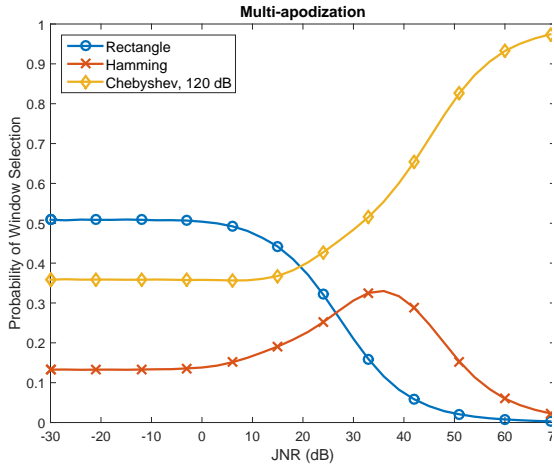}
        \caption{Multi-apodization Method}
    \end{subfigure}%
    \begin{subfigure}[t]{0.5\textwidth}
        \centering
\includegraphics[height=6.5cm]{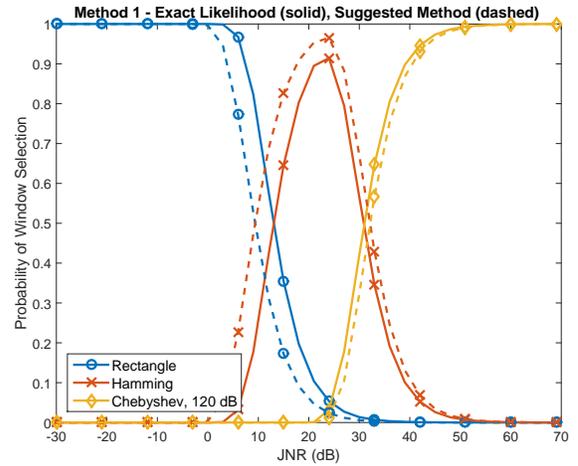}
        \caption{Proposed Methods}
    \end{subfigure}
   \caption{ Case 1. Subfigures (a)-(b): ROC curves for the windowed DFT detectors, multi-apodization method and proposed detectors, $P_{FA} = 10^{-2}$. Subfigures (c)-(d): The window selection probabilities of multi-apodization and proposed methods vs. JNR.}
\label{fig_case1}
\end{figure}

\begin{figure}[t!]
    \centering
    \begin{subfigure}[t]{0.5\textwidth}
        \centering
\includegraphics[height=6.5cm]{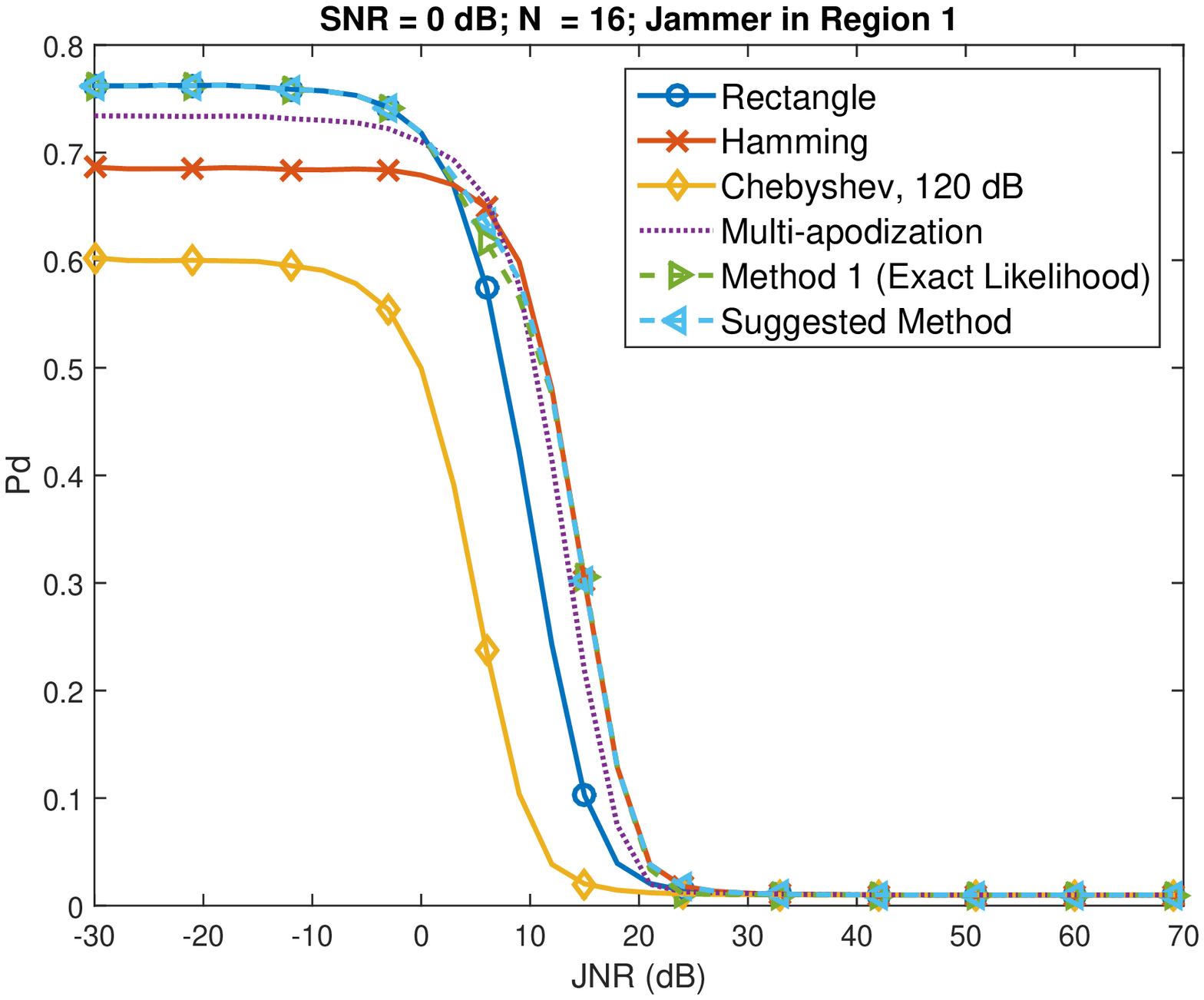}
        \caption{Pd vs. JNR}
    \end{subfigure}%
    \begin{subfigure}[t]{0.5\textwidth}
        \centering
\includegraphics[height=6.5cm]{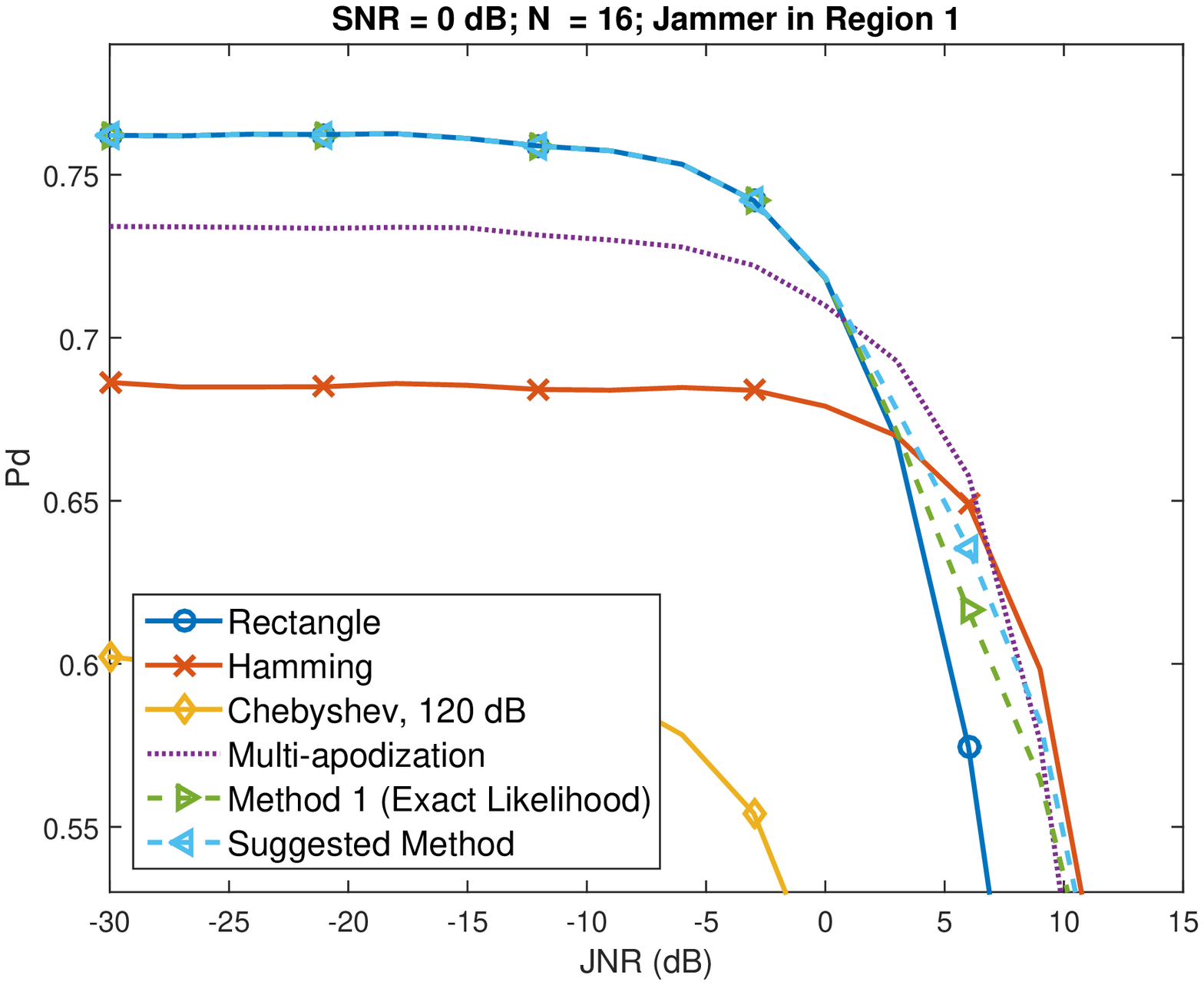}
        \caption{Pd vs. JNR (zoomed)}
    \end{subfigure}

\begin{subfigure}[t]{0.5\textwidth}
        \centering
\includegraphics[height=6.5cm]{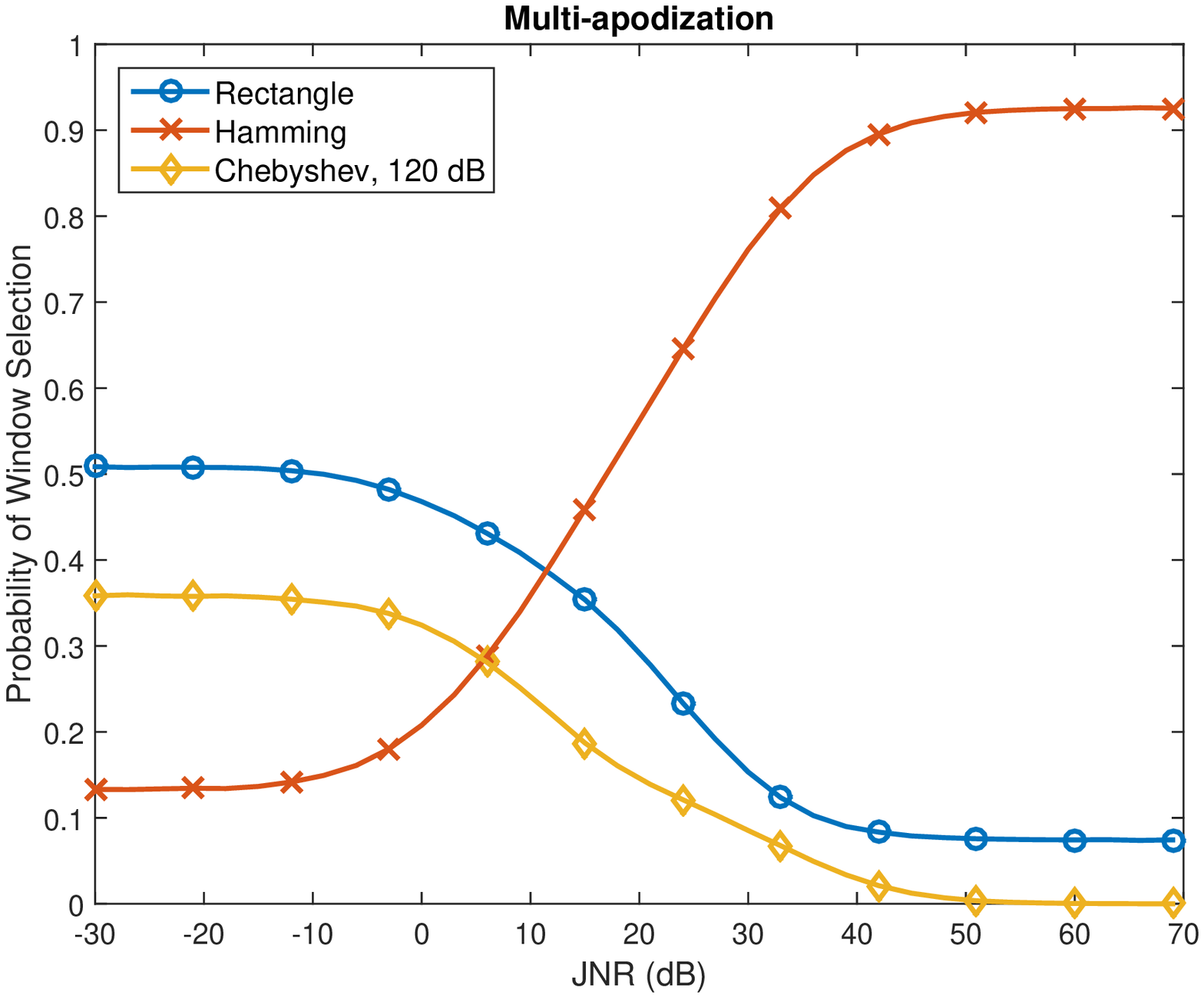}
        \caption{Multi-apodization Method}
    \end{subfigure}%
    \begin{subfigure}[t]{0.5\textwidth}
        \centering
\includegraphics[height=6.5cm]{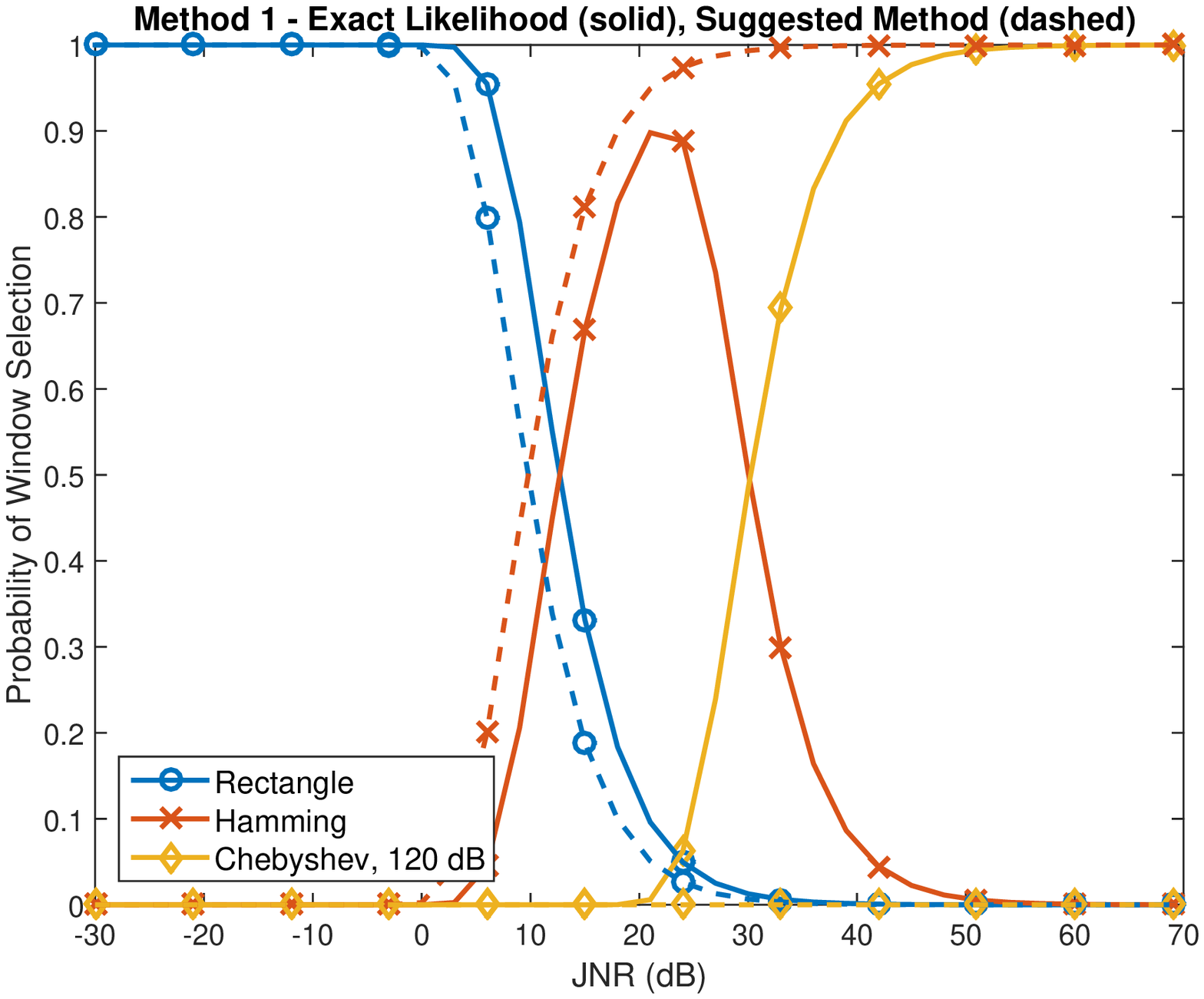}
        \caption{Proposed Methods}
    \end{subfigure}
   \caption{ Case 2. Subfigures (a)-(b): ROC curves for the windowed DFT detectors, multi-apodization method and proposed detectors, $P_{FA} = 10^{-2}$. Subfigures (c)-(d): The window selection probabilities of multi-apodization and proposed methods vs. JNR.}
\label{fig_case2}
\end{figure}

Figures~\ref{fig_case1}a-\ref{fig_case1}b show the variation of the probability of detection versus jammer-to-noise ratio (JNR) for the scenario parameters of SNR  = 0 dB, the probability of false alarm of $10^{-2}$, the number of observations of $16$, ($N=16$).  The frequency of the interfering signal is 4 to 6 DFT bins away from the frequency of the signal to be detected and  the interfering frequency is uniformly distributed in this interval guaranteeing that the interference is in Region 2 of Figure~\ref{fig_windows}.

From Figures~\ref{fig_case1}a-\ref{fig_case1}b, it can be immediately read that the rectangle window presents the best performance (in the sense of detection probability) for sufficiently low JNR values. On the other hand, the performance of rectangle windowed Fourier transformation detector degrades rapidly, once JNR is above $20$ dB. It should be clear from these figures that the windows providing higher interference suppression capabilities should be utilized at high JNR levels.

In many practical systems, a nominal window function, such as the Hamming window, is selected by the system designer and this window is utilized irrespective of the JNR value. Figure~\ref{fig_case1}a-\ref{fig_case1}b illustrate that the choice of Hamming window brings a sub-optimal performance at low JNR and furthermore does not present a significant gain at excessively high JNR values. Typically, the system designer rules out the possibility of excessive JNR values through another jammer detection mechanism, say a side-lobe blanker, and justifies the SNR loss due to the application of Hamming window as a necessary trade-off between low JNR and medium JNR case. This study aims to present an alternative method based on a data-adaptive window selection procedure in which SNR loss due to windowing is effectively eliminated.

\begin{figure}[t!]
    \centering
    \begin{subfigure}[t]{0.5\textwidth}
        \centering
\includegraphics[height=6.5cm]{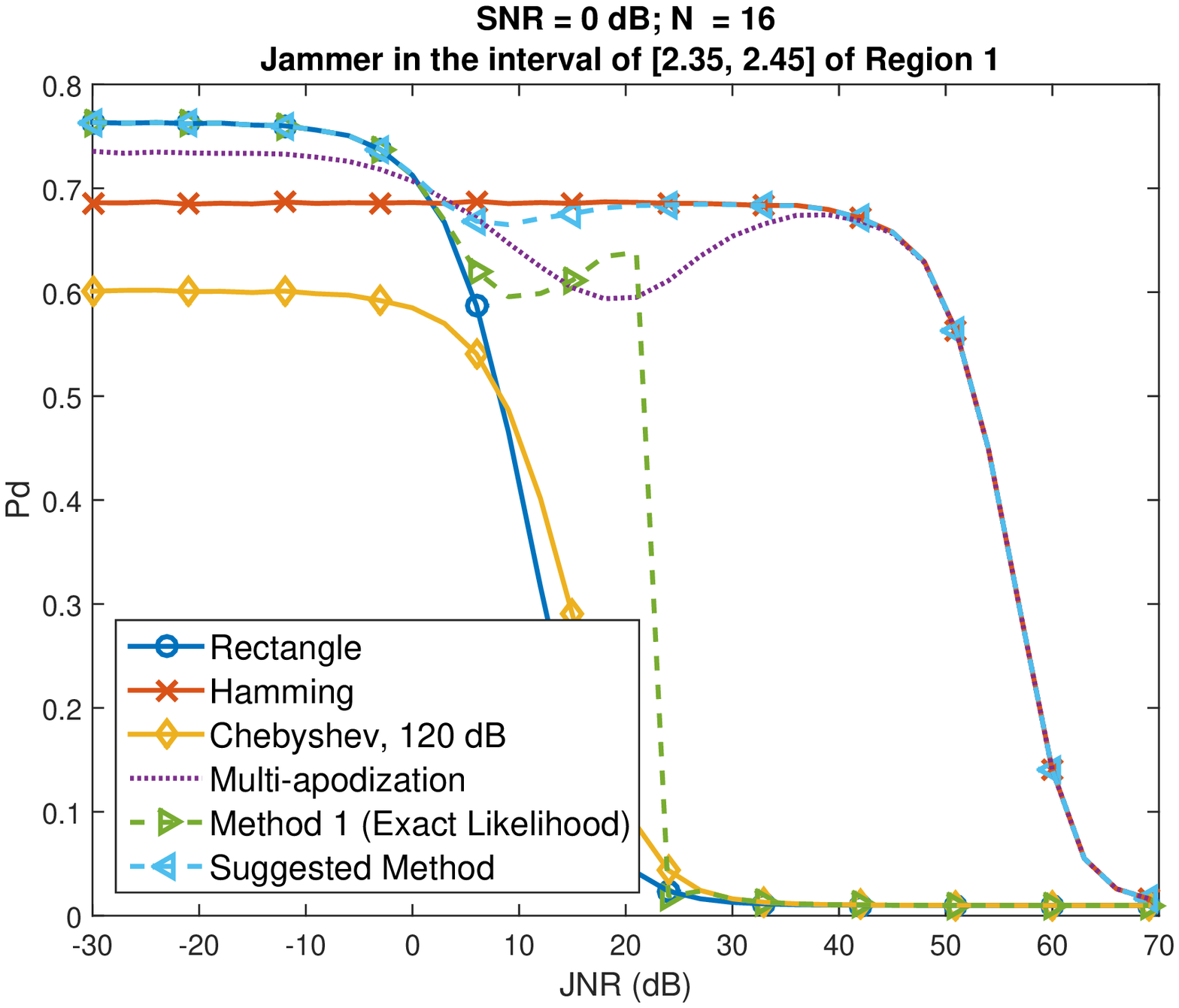}
        \caption{Pd vs. JNR}
    \end{subfigure}%
    \begin{subfigure}[t]{0.5\textwidth}
        \centering
\includegraphics[height=6.5cm]{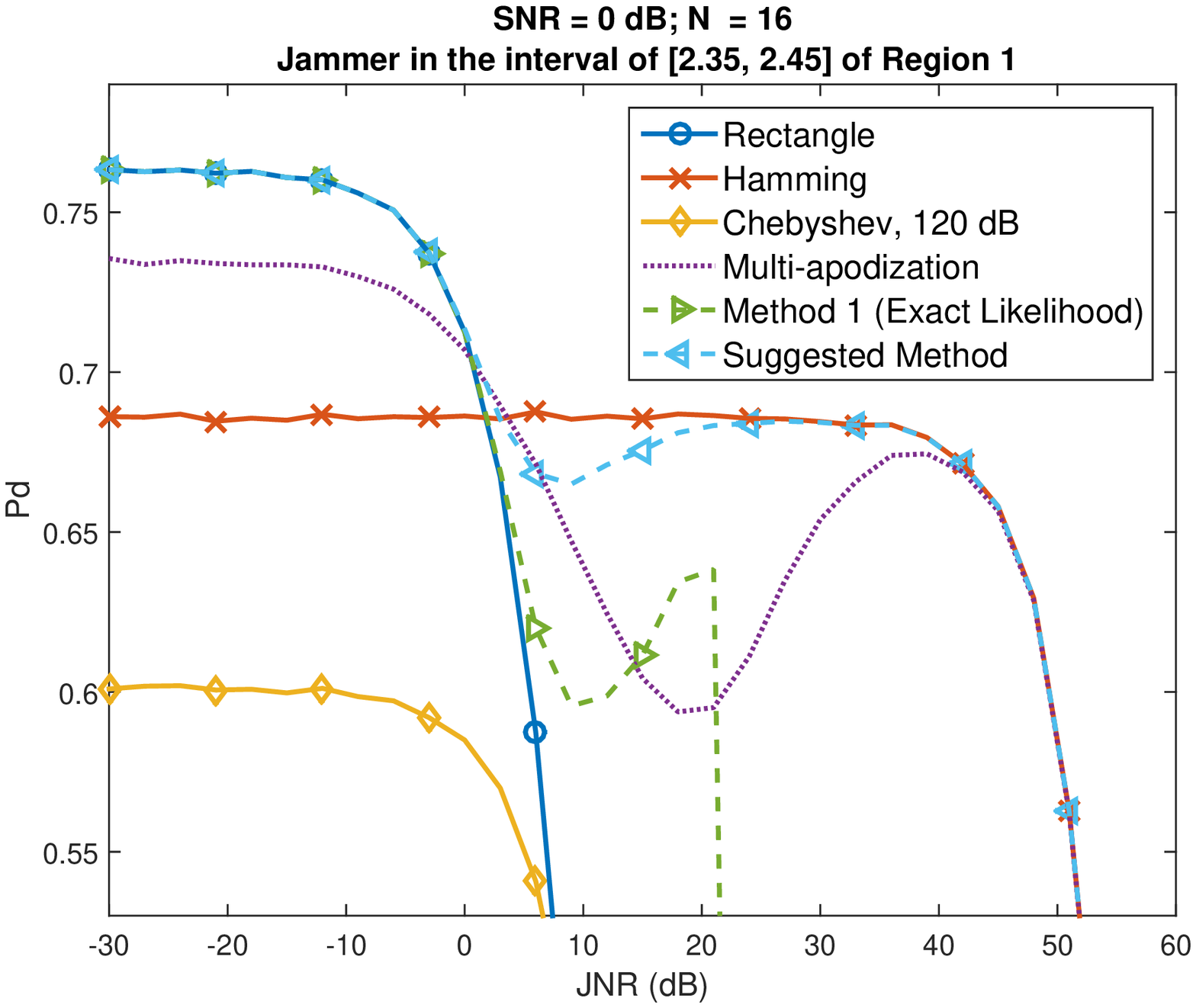}
        \caption{Pd vs. JNR (zoomed)}
    \end{subfigure}

\begin{subfigure}[t]{0.5\textwidth}
        \centering
\includegraphics[height=6.5cm]{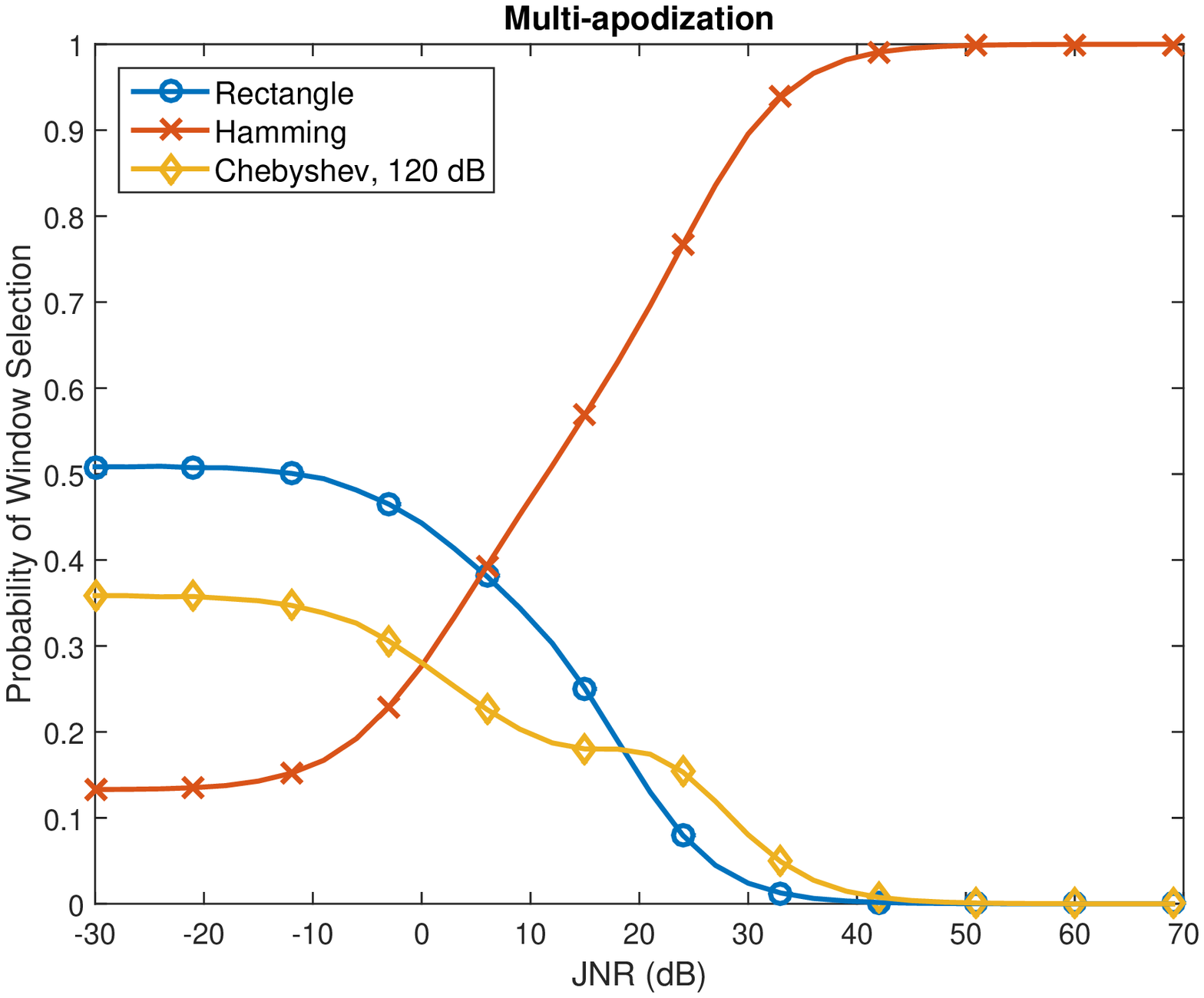}
        \caption{Multi-apodization Method}
    \end{subfigure}%
    \begin{subfigure}[t]{0.5\textwidth}
        \centering
\includegraphics[height=6.5cm]{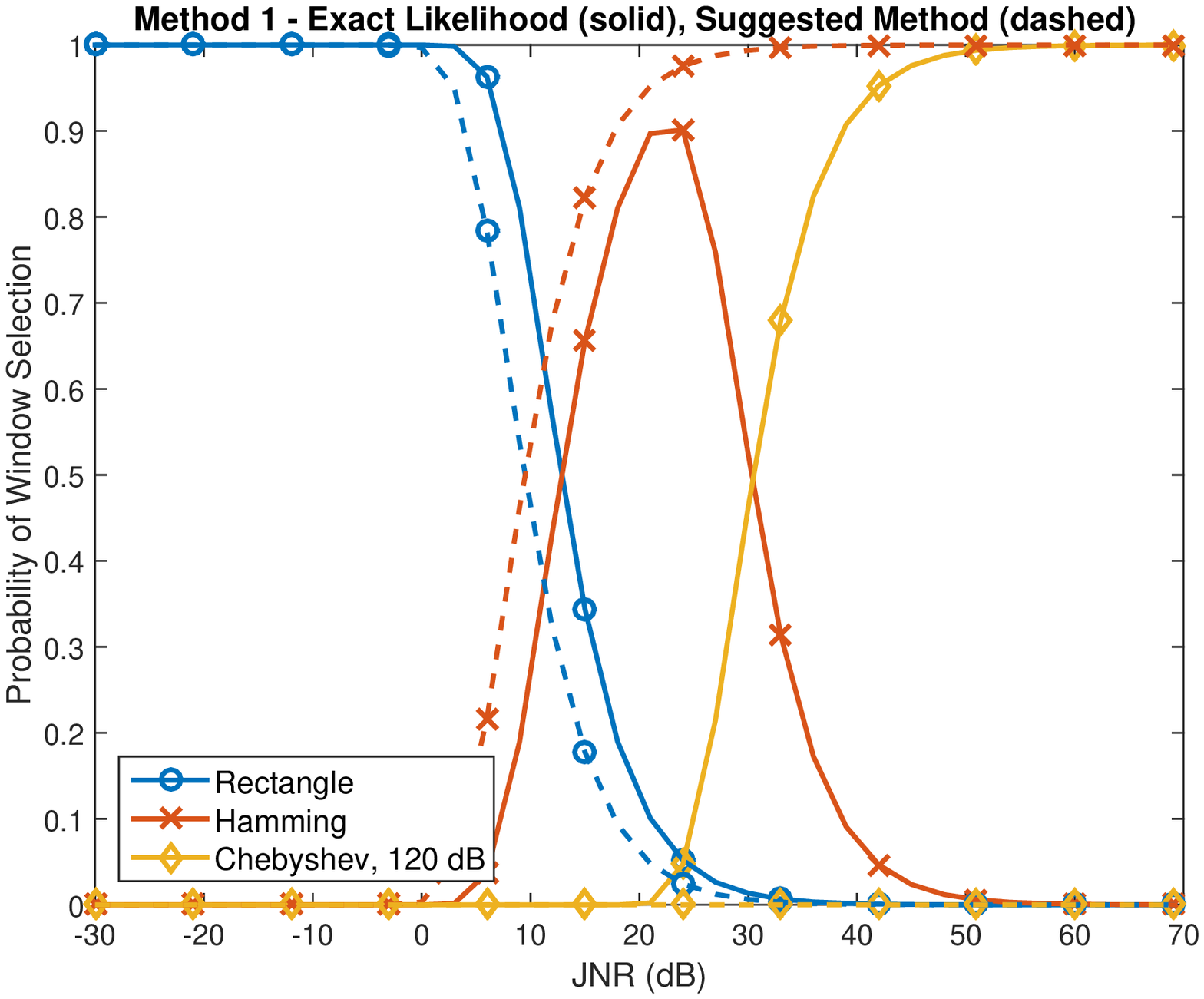}
        \caption{Proposed Methods}
    \end{subfigure}
   \caption{Case 3. Subfigures (a)-(b): ROC curves for the windowed DFT detectors, multi-apodization method and proposed detectors, $P_{FA} = 10^{-2}$. Subfigures (c)-(d): The window selection probabilities of multi-apodization and proposed methods vs. JNR.}
\label{fig_case3}
\end{figure}

The dashed line in Figure~\ref{fig_case1}a-\ref{fig_case1}b shows the performance of the multi-apodization method. The window selection procedure of this method is established through the relation given in \mref{eq2}. This method can be summarized as follows. The windowed DFT of the input is calculated for all three windows. For a fixed DFT bin of interest, the windowed DFT output with the least magnitude is selected. (It is important to note that the windowed DFT outputs should have the same response when the incoming signal frequency perfectly matches the DFT bin frequency in the absence of noise. This is easily established by equating the DC response of the windows to 1 (equivalently 0 dB) by scaling each window, as shown in Figure~\ref{fig_windows}.) When the dashed line in Figure~\ref{fig_case1}a-\ref{fig_case1}b (multi-apodization method) is compared with the solid lines (conventional windowed DFT detectors), we see that at low JNR values, the multi-apodization method yields a better performance than both Hamming and Chebyshev windows. Figure~\ref{fig_case1}c shows, the probability of window selection for the multi-apodization method. This figure shows that at low JNR values, about 50\% of the time rectangle window is selected. This explains the superiority of multi-apodization method over Hamming and Chebyshev windows at low JNR. Similarly, when JNR is extremely high, say 70 dB, the performance of the multi-apodization method approaches Chebyshev window, which is the best choice among the windows examined for excessively high JNR values. Figure~\ref{fig_case1}c illustrates that as JNR increases, the probability of Chebyshev window selection also increases, explaining high JNR behavior of multi-apodization method in Figure~\ref{fig_case1}a-\ref{fig_case1}b. For medium-to-high JNR values, the multi-apodization method presents an acceptable performance, as seen from Figure~\ref{fig_case1}b. The main of reason of less-than-impressive performance is the sluggish change of window selection probabilities as shown in Figure~\ref{fig_case1}c. Stated differently, the multi-apodization method does not react fast enough to JNR increase. 

The performance of the proposed method with exact and approximate likelihood metric is shown with the dashed lines and triangle markers in Figure~\ref{fig_case1}a/b. It can be immediately noted that the performance with the exact and approximate likelihood metric are almost identical. The performance of suggested method is superior to the windowed based approach at all JNR values and is almost identical to the pointwise maximum of three detectors shown in Figure~\ref{fig_case1}a-\ref{fig_case1}b at all JNR levels. The superior performance in the detection probability can also be explained by the rapid variation of the window selection probabilities given in Figure~\ref{fig_case1}d.

\textbf{Case 2: Interference in the transition band of a window:}
Here we assume that interference is in the transition band of one of the windows.
For this case, the frequency of the interfering signal is assumed to be uniformly distributed 1.5 to 3 DFT bins away from the frequency of the signal to be detected, i.e. in Region 1 of Figure~\ref{fig_windows}. The other simulation parameters are identical to ones in Case~1.

Figure~\ref{fig_case2} presents the results for this scenario. It can be noted the multi-apodization detector and proposed detector approach the performance of Hamming window as JNR increases. From Figure~\ref{fig_case2}c, it can be noticed that the multi-apodization output is dominated by the selection of the Hamming window at high JNR.
Similarly, as the dashed red line with the cross marker indicates the suggested method also utilizes Hamming window with increasing JNR, but the proposed method adopts the Hamming window at much smaller JNR values in comparison to the multi-apodization method.
The Chebyshev window is not selected at all with the proposed method. It should be intuitively clear that the choice of Chebyshev window should be avoided for this scenario. The next case further elaborates this point.
\begin{sidewaystable}[t]
\caption{Rectangle (R), Hamming (H) and Chebyshev (C) window selection probabilities for each DFT bin when input is the superposition of two complex exponential signals of length $N=16$ with the frequencies $\frac{2\pi}{N}0.1$ and $\frac{2\pi}{N}6.25$ radians per sample, i.e. frequencies of 0.1 and 6.25 with the units of DFT bins.}
\begin{tabular}{|l|l|llllllllllllllll|}
\hline
& & \multicolumn{16}{c|}{DFT Bin Index} \\
\hline
\  & \ & 0 & 1 & 2 & 3 & 4 & 5 & 6 & 7 & 8 & 9 & 10 & 11 & 12 & 13 & 14 & 15 \\
\hline \hline
\ \ \ \textbf{Case A} & R &
85.2 & 84.8 & 79.9 & 79.6 & 80.2 & 97.2 & 99.7 & 99.8 & 84.3 & 79.7 & 79 & 79.1 & 79.1 & 79.6 & 79.4 & 84.1 \\
SNR$_1$=0 dB, & H &
14.8 & 15.2 & 20.1 & 20.4 & 19.8 & 2.8 & 0.3 & 0.2 & 15.7 & 20.3 & 21 & 20.9 & 20.9 & 20.4 & 20.6 & 15.9 \\
SNR$_2$=5 dB & C &
0 & 0 & 0 & 0 & 0 & 0 & 0 & 0 & 0 & 0 & 0 & 0 & 0 & 0 & 0 & 0 \\
\hline \hline
\ \ \ \textbf{Case B} & R &
17.8 & 17.7 & 15.4 & 15.2 & 15.5 & 33.6 & 86.6 & 80 & 17.6 & 15.5 & 15.4 & 15.4 & 15.3 & 15.4 & 15.3 & 17.5 \\
SNR$_1$=0 dB, & H &
82.2 & 82.3 & 84.6 & 84.8 & 84.5 & 66.4 & 13.4 & 20 & 82.4 & 84.5 & 84.6 & 84.6 & 84.7 & 84.6 & 84.7 & 82.5 \\
SNR$_2$=15 dB & C &
0 & 0 & 0 & 0 & 0 & 0 & 0 & 0 & 0 & 0 & 0 & 0 & 0 & 0 & 0 & 0 \\
\hline \hline
\ \ \ \textbf{Case C} & R &
1.8 & 1.8 & 1.6 & 1.6 & 1.7 & 3.9 & 18.8 & 15.7 & 1.9 & 1.6 & 1.6 & 1.6 & 1.5 & 1.5 & 1.5 & 1.7 \\
SNR$_1$=0 dB, & H &
95.7 & 95.7 & 96 & 96 & 98.3 & 96.1 & 81.2 & 84.3 & 98.1 & 98.4 & 96 & 96 & 96 & 96 & 96 & 95.8 \\
SNR$_2$=25 dB & C &
2.5 & 2.5 & 2.4 & 2.4 & 0 & 0 & 0 & 0 & 0 & 0 & 2.4 & 2.4 & 2.5 & 2.5 & 2.5 & 2.5 \\
\hline \hline
\ \ \ \textbf{Case D} & R &
0.4 & 0.4 & 0.1 & 0.1 & 0.1 & 0.6 & 2.1 & 1.9 & 0.2 & 0.1 & 0.1 & 0.1 & 0.1 & 0.1 & 0.1 & 0.4 \\
SNR$_1$=0 dB, & H &
30.4 & 30.5 & 30.9 & 31.2 & 99.9 & 99.4 & 97.9 & 98.1 & 99.8 & 99.9 & 31 & 30.9 & 30.7 & 30.6 & 30.6 & 30.3 \\
SNR$_2$=35 dB & C &
69.2 & 69.1 & 69 & 68.7 & 0 & 0 & 0 & 0 & 0 & 0 & 68.9 & 69 & 69.2 & 69.3 & 69.3 & 69.3 \\
\hline \hline
\end{tabular}
\label{table_win_select}
\end{sidewaystable}

\textbf{Case 3: Interference in the transition band of a window:}
Figure~\ref{fig_case3} presents the results when the interfering frequency is 2.35 to 2.45 bins away from the frequency of the signal to be detected. The interval for the interfering frequency is specially selected to illustrate the disadvantages arising when the interference is in the transition band of a window. From Figure~\ref{fig_windows}, it can be noted that Hamming window presents around 80 dB in the interval of [2.35,2.45] bins; while the Chebyshev window presents a suppression ration of 20 dB. Ideally, the Hamming window should be the most suitable choice for the suppression of a strong interferer in this frequency interval. It should be noted this situation, i.e. the preference of a window having a poorer peak-sidelobe suppression ratio, occurs only when the interfering frequency is in the transition band of one of the windows.



Figure~\ref{fig_case3} presents the performance of suggested method. The dashed lines in Fig~\ref{fig_case3}d indicate that the Chebyshev window is not selected at all JNR values with the suggested method. It can be said that the suggested method, in effect, eliminates the Chebyshev window choice when the interference is in the transition band of this window as desired. It can also be seen from Figures~\ref{fig_case3}a-\ref{fig_case3}b that high JNR performance of the suggested method is identical to the performance of the Hamming window, as desired. We also note that the method utilizing the exact likelihood (which is not equipped by window disabling feature) only works well when the interferer is in the stop-band of all window functions, as in Case 1.

\textbf{Study of Window Selection Probability:} As a final numerical study, we examine the probability of window selection in more depth. The input for this study is assumed to be
\be
r[n] = \sqrt{\gamma_1}e^{j(\omega_1 n + \phi_s)} + \sqrt{\gamma_2}e^{j(\omega_2 n + \phi_j)} + v[n], \quad n=\{0, 1, \ldots, 15\} \nn
\ee
where $\omega_1 = \frac{2\pi}{16}0.1$, $\omega_2 = \frac{2\pi}{16}6.25$  and $v[n]$ is unit variance, zero-mean white noise. We continue to use the SNR definition in Section~\ref{sec:background}. Under this setting, Table~\ref{table_win_select} shows the window selection probabilities for the proposed method when signal components forming $r[n]$ has different SNR values.

It can be immediately observed that the window selection probabilities vary with the DFT bin of interest. Stated differently, if we are interested in the frequency content of the 11th DFT bin, both components forming $r[n]$, i.e. signals at the bins 0.1 and 6.25, act like jammers corrupting the frequency content of this bin.

Case A of Table~\ref{table_win_select} illustrates the case for weak signal components. It can be noticed that the rectangle window (whose results are shown in the rows indicated by 'R' letter) is selected with a significant majority for the frequency bins close to the signal components (0'th, 6'th and 7'th bins). The Hamming window is selected 14.8\% of the time for the DC bin. This is essentially due to interference generated by the signal component at the 6.25 DFT bin. The selection of the Hamming window for 6th or 7th DFT bins is much lower, close to 0.3\%, since the signal at DC bin imposes has a 5 dB lower SNR and therefore causes less interference. For the weak jamming signal case, it is assuring to observe that the window with the highest SNR loss, i.e. the Chebyshev window, is not utilized at all.

Case B of Table~\ref{table_win_select} gives the window selection probabilities when the second component with the spectral location of 6.25 DFT bin has an SNR of 15 dB. It can be observed that the weak signal having the frequency of 0.1 DFT bins is processed with the Hamming window 82.8\% of the time. The percentage increase from 14.8\% (Case A) to 82.8\% (Case B) is essentially due to power increase of the second component. One can also notice that the Hamming window utilization for the 6th bin also increases from 0.3\% to 13.4\% when Cases A and B are compared. The power of the first signal component is identical in both cases; hence this increase can not be explained with the increase of jamming activity due to the first component. The increase in the Hamming window utilization probability is due to the frequency mismatch of the second component (6.25 DFT bin) to frequencies of DFT bins, i.e. integer valued bins. Due to the mentioned frequency mismatch, the signal energy in the side-lobes of this signal acts as an interference, inhibiting its detection. This problem is known as the parameter mismatch problem (the mismatch of assumed frequencies with the actual signal frequency) and can be reduced by evaluating zero-padded DFT instead of N-point DFT. We do not further elaborate on this topic in order not to distract the readers from the main message of this study.

Cases C and D of Table~\ref{table_win_select} present the results as the second signal component SNR is further increased. Only in the highest $\textrm{SNR}_2$ case (Case D), the window preference switches to the Chebyshev window. It should be noted that this change of preference occurs for the bins that the second signal component lies in the stop-band of the Chebyshev window. Stated differently, the bins 4 to 10 do not use Chebyshev at all, since these bins constitute the main lobe and transition region of this window. Hence, for these bins the Chebyshev window can not provide any interference suppression capability. A brief of time of reflection on Table~\ref{table_win_select} can convince the readers that the suggested method works as if like an experienced operator with the knowledge of jammer specific parameters.

\section{Conclusions}
We present a method without any application specific parameter tuning that automates the window selection procedure for the discrete Fourier transformation (DFT) based detectors. The windowing is, in general, an essential operation to reduce the masking effect of a strong spectral component over the weaker one. Yet, it may come at a significant SNR loss, typically in between 1.5 to 3 dB. The suggested method aims to select the windows with strong side-lobe suppression capabilities, or equivalently the windows with a high SNR loss, only when the situation arises, that is only when it is indeed needed. The numerical results show that the method is capable of switching window function depending on operational JNR level such that the best probability of detection among all detectors is achieved at all JNR values, in spite of the lack of JNR knowledge at the receiving end (see Figures~\ref{fig_case1} to \ref{fig_case3}). We believe that the suggested method is a promising contender for the open title of single snapshot version of the Capon's method. An implementation of the method is given in \cite{ccandan-code}.


\bibliography{windowing_with_no_loss}
\end{document}